\documentclass[conference]{IEEEtran}
\usepackage[T1]{fontenc}

\usepackage[latin5]{inputenc}
\usepackage{times}
\usepackage[figuresright]{rotating}
\usepackage{float}
\usepackage{subcaption}
\usepackage{url}
\usepackage{setspace}
\usepackage{balance}
\usepackage{multirow}
\usepackage{color}
\usepackage{graphicx}
\usepackage[implicit=false]{hyperref}
\usepackage{epsfig}
\usepackage{amsmath, amssymb}

\usepackage{capt-of}
\usepackage{multirow}
\usepackage{array}
\usepackage{caption} 
\usepackage{pslatex}
\usepackage{enumitem}

\usepackage[dvipsnames]{xcolor}
\usepackage{tcolorbox}
\usepackage{tabularx}
\tcbset{tab2/.style={colback=SkyBlue!5!white,colframe=blue!50!black,colbacktitle=Blue!40!white,
coltitle=black,center title}}

\usepackage{colortbl}
\makeatletter
\newcolumntype{"}{@{\hskip\tabcolsep\vrule width 1pt\hskip\tabcolsep}}
\makeatother
\newcolumntype{?}{!{\vrule width 1.5pt}}

\makeatletter
\newcommand{\thickhline}{%
    \noalign {\ifnum 0=`}\fi \hrule height 1.5pt
    \futurelet \reserved@a \@xhline
}
\newcolumntype{"}{@{\hskip\tabcolsep\vrule width 1pt\hskip\tabcolsep}}
\makeatother
\newcolumntype{?}{!{\vrule width 1.5pt}}
\ifCLASSOPTIONcompsoc
  \usepackage[nocompress]{cite}
  \else
  \usepackage{cite}
\fi
\ifCLASSINFOpdf
\else
\fi
\begin{document}

\newcommand{\squeezeupppp}{\vspace{-8 mm}}
\newcommand{\squeezeuppp}{\vspace{-6 mm}}
\newcommand{\squeezeupp}{\vspace{-5 mm}}
\newcommand{\squeezeup}{\vspace{-3 mm}}
\newcommand{\squeezeu}{\vspace{-2 mm}}
\newcommand{\squeeze}{\vspace{-1 mm}}
%
\title{Fieldable Cross-Layer Optimized Embedded Software Defined Radio is Finally Here! 
}
%
%
%
%

\author{\IEEEauthorblockN{Jithin Jagannath, Anu Jagannath, Justin Henney, Noor Biswas, Tyler Gwin, Zackary Kane, Andrew Drozd}
\IEEEauthorblockA{Marconi-Rosenblatt AI/ML Innovation Lab, ANDRO Computational Solutions, LLC, Rome NY \\ \{jjagannath, ajagannath, jhenney, nbiswas, tgwin, zkane, adrozd\}@androcs.com
\squeezeupp
}}

\maketitle


\begin{abstract}
The concept of cross-layer optimization has been around for several years now. The primary goal of the cross-layer approach was to liberate the strict boundary between the layers of the traditional OSI protocol stack. This is to enable information flow between layers which then can be leveraged to optimize the network's performance across the layers. This concept has been of keen interest for tactical application as there is an overwhelming requirement to operate in a challenging and dynamic environment. The advent of software defined radios (SDR) accelerated the growth of this domain due to the added flexibility provided by SDRs. Even with the immense interest and progress in this area of research, there has been a gaping abyss between solutions designed in theory and ones deployed in practice. To the best of our knowledge, this is the first time in literature, an embedded SDR has been leveraged to successfully design a cross-layer optimized transceiver that provides high throughput and high reliability in a ruggedized, weatherized, and fieldable form-factor. The design ethos focuses on efficiency and flexibility such that optimization objectives, cross-layer interactions can be reconfigured rapidly. To demonstrate our claims, we provide results from extensive outdoor over-the-air evaluation in various settings with up to 10-node network typologies. The results demonstrate high reliability, throughput, and dynamic routing capability achieving high technology readiness level (TRL) for tactical applications.   
\end{abstract}




%
\IEEEpeerreviewmaketitle

\section{Introduction}\label{sec:introduction}

Wireless communication has undoubtedly become a ubiquitous part of our lives and we are constantly striving  to meet the evolving needs. This includes all connected devices in our smart homes, our cellular network, the entire concept of internet-of-things (IoT) networks controlling manufacturing, industry automation, smart grid metering, space communications, underwater networks, tactical networks among others. As we move from 5G (5th Generation) to 6G (6th Generation), the need to optimize the scarce resources is becoming evident and inevitable \cite{6Genesis,6Gcomms,jagannath2020redefining, 6gsaad,6Gfrontier}. 

Traditionally, the strictly layered architecture proposed by the open systems interconnection (OSI) reference model has been the prevalent design for a majority if not all modern networking architectures. This is strict in the sense that they are designed to maintain only a limited interface between the neighboring layers \cite{Fu_survey}. Realizing the deficiencies in this layered architecture, cross-layer optimized approach has been proposed over the past decade to maximize the utilization of scarce resources by "erasing" the strict boundaries between various layers of the protocol stack. In other words, any attempt to violate the OSI reference model is considered a cross-layer design \cite{Foukalas_survey}. While there are abundant solutions proposed in literature \cite{Fu_survey}, the majority of it is limited to simulations that may have strong assumptions and/or do not consider all the hardware constraints and rigidness that may be encountered during a real-life deployment. During the next phase of advancement, the advent of software defined radios (SDR) provided the much-needed impetus to this concept providing the flexibility to implement novel cross-layer architectures. This enabled some of these efforts to be extended to hardware-based testbed evaluations. In most cases, these efforts still used one or more dedicated (non-embedded) host computers to implement the solutions which were then connected to SDRs. Even with these advances, to the best of our knowledge, there does not exist a ruggedized fieldable SDR with a comprehensive cross-layer optimization capable software module implemented on an embedded ARM processor. The main reason for this is the various hurdles that are associated with developing the solution from theory to effective hardware deployable software. 

\textbf{Contribution:} In this article, we present the first, completely stand-alone, ruggedized, and fieldable cross-layer optimized solution build using a low SWaP (Size, Weight, and Power) SDR. In this case, we implement an energy-aware cross-layer protocol that aims to maximize network lifetime for enabling telemetry collection of tactical test and evaluation ranges. The primary contribution of this work is the realization of theoretical or simulation-level concepts to a fieldable hardware entity. We first discuss the hardware-level modification required to customize the baseline SDR into a fieldable solution. Next, we outline how the protocol stack was designed and implemented on a computationally constrained ARM processor. To demonstrate the feasibility of the implementation and the designed transceiver in terms of throughput and reliability, we performed extensive outdoor experiments with up to 10 nodes in the network.

\textbf{Impact:} In this paper, we have demonstrated the feasibility and effectiveness of designing and developing an embedded SDR-based cross-layer optimized solution. The proposed architecture and design principles can be leveraged to implement and mature several of the novel cross-layer optimized solutions to meet the evolving needs of both tactical and commercial communication systems. Our implementation principle has been to facilitate rapid reconfiguration of network objective with only few lines of code thereby enabling a \emph{truly} software-defined radio that evolves with the growing requirements. 
We hope and believe the unique cross-layer design methodologies and extensive outdoor evaluations (field trials) will serve as an impetus for maturation of novel cross-layer solutions.


\section{Related Work}

Cross-layer approaches have been explored as a novel solution to address a wide range of problems in wireless communication due to its perceived benefits in sustaining communications in a dynamic and constrained environment \cite{Fu_survey, Foukalas_survey}. This includes RF domain for tactical network \cite{Wang_cross, Nosheen_cross_OMNET, DRS}, commercial network \cite{Zhu_TCP_cross, Herrmann_energy_cross, Jagannath19CCNC_1, Barmpounakis_crosslayer_5G}, acoustic underwater networks, and even in the upcoming visible light communication networks \cite{Zhou_cross, Nancy18ADH, Demir_VLC, Jagannath19WOWMoM}. These solutions are designed for various objectives such as optimizing throughput and/or latency \cite{ROCH, ROSA, She_URLLC, Jagannath18TMC}, fairness \cite{Wang_cross}, energy consumption \cite{Herrmann_energy_cross, Jagannath19CCNC_1, Zhou_cross}, resource management \cite{Barmpounakis_crosslayer_5G, Demir_VLC}, efficient multi-path TCP \cite{Zhu_TCP_cross}. In this work, we focus on reviewing the maturity of these state-of-the-art cross-layer solutions to \textit{expose the absence of deployable embedded SDR based solutions}. 

As in any novel research, simulations are the obvious first choice to establish feasibility and performance gains over existing approaches. Most works rely on MATLAB \cite{ROSA, DRS}, NS3 \cite{Herrmann_energy_cross}, OMNET \cite{Nosheen_cross_OMNET} or similar simulators. Majority of the works that propose cross-layer optimization have been limited to simulation \cite{ROSA,DRS,Herrmann_energy_cross, Wang_cross,Nosheen_cross_OMNET,Zhu_TCP_cross,ROCH, She_URLLC} due to the challenges, time, and effort it takes to evaluate them on hardware testbeds. In most cases, these simulations are executed under various assumptions and/or are abstracted from the physical layer (PHY) of the protocol stack and are restricted to packet-level simulators \cite{DRS, ROSA, Fu_survey}. This abstraction implies several intricacies of real-world deployment are overlooked or set aside to be handled in the future which often thwarts the maturation level of cross-layer solutions.

In several cases, the next logical step is to evaluate the feasibility and performance on hardware-based testbed. The advent of SDR has significantly bolstered efforts in this direction. At the same time, there is a distinction in the maturity/utility of solutions that have been implemented using a host PC controlling the SDRs and ones efficiently ported to a standalone embedded SDR. The key difference is in the rapidity of development on computationally capable hardware as opposed to carefully optimized (often C/C++ or VHDL/Verilog) implementation on embedded (resource-constrained) hardware. 
The host PC based development approach saves time and resources to rule out impractical solutions before significant time is spend in optimizing the implementation for final mature deployment. Therefore, based on the resources, the risk associated with novel cross-layer solutions either approaches can be adopted to mature network optimizing solutions. Several of the solutions discussed earlier have been successfully extended to preliminary hardware testbeds \cite{Sklivanitis15GLOBECOM,Jagannath18TMC}. In most of these cases, SDRs like the USRPs are used in association with the host PC. Beyond SDRs, well-defined commercial wireless protocols like WiFi and LoRa (for PHY) have also been used along with microcomputing platforms to design cross-layer approaches \cite{Jagannath19ADH_HELPER}. In this case, a cross-layer approach - distributed energy-efficient routing (SEEK) - was implemented on  Raspberry Pi to maximize the network lifetime using LoRa as the PHY. In this form, it was highly restricted in throughput (due to LoRa) as a trade-off for longer transmission range.  

To summarize, the majority of the cross-layer optimized work is limited to simulations and has not been successfully validated on a hardware platform. This is a major hurdle and shortcoming of the current state of research and development. In the recent past, some of this has been mitigated by preliminary hardware-based evaluations but has often provided limited performance in terms of metrics like throughput (due to low sampling rate constrained by computations) and/or has been depended on external computational platforms. Due to these reasons, even with the advances made in the field, there is no known cross-layer optimized embedded SDR solution built using commercial-off-the-shelf (COTS) embedded SDR that, (i) can be deployed as a standalone unit, (ii) makes cross-layer optimized distributed routing decisions, (iii) is ruggedized for outdoor deployment, and (iv) can provide reliable and high throughput (up to 11 Mbps) links over large distances (1 km for up to 5.5 Mbps). 

\section{System Design}
\label{Sec:SystemDesign}

In this section, we describe the system design of the cross-layer optimized transceiver and how it has been executed.

\subsection{Embedded Software Defined Radio Platform}


One of the key objectives of the work was to ensure a modular and programmable, portable, handheld, battery-powered standalone solution which should operate in harsh conditions for several hours. This implied that the foundation of the design needs to be a low SWaP embedded SDR. The proposed solution was implemented on a Epiq Solutions' Sidekiq Z2 SDR \cite{sidekiqz2} (Fig. \ref{fig:Z2}). It consists of an Analog Devices' AD9364, Xilinx Zynq XC7Z010-2I system on chip (SoC), and the key device specifications are provided in Table. \ref{tab:z2}

\begin{center}
\begin{minipage}[h]{0.64 \linewidth}
 \small
 \centering
 \captionof{table}{Specification of Z2}
 \squeezeup
    \begin{tcolorbox}[tab2,tabularx={|p{1.7 cm}||p{3 cm}}]
      \textbf{Specs.}   & \textbf{Values}  \\ \hline\hline
      Frequency   & 70 MHz - 6 GHz  \\ \hline
    Sample Rate & Up to 61.44 MS/s   \\ \hline
    Size & 30 x 51 x 5 mm \\ \hline
    Weight & 8 grams \\ \hline
    Processor & Dual-core ARM \\ \hline
    \end{tcolorbox}{}
    
    \label{tab:z2}
\end{minipage}
\begin{minipage}[h]{0.34 \linewidth}
\centering
\captionof{figure}{Z2 SDR}
\squeezeup
\includegraphics[width=.6 \columnwidth]{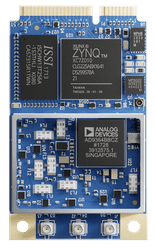} 
\label{fig:Z2}
\end{minipage}
\end{center}

\begin{figure*}[h!]
\begin{minipage}[h]{0.26 \linewidth}
\centering
\includegraphics[width=.99 \columnwidth]{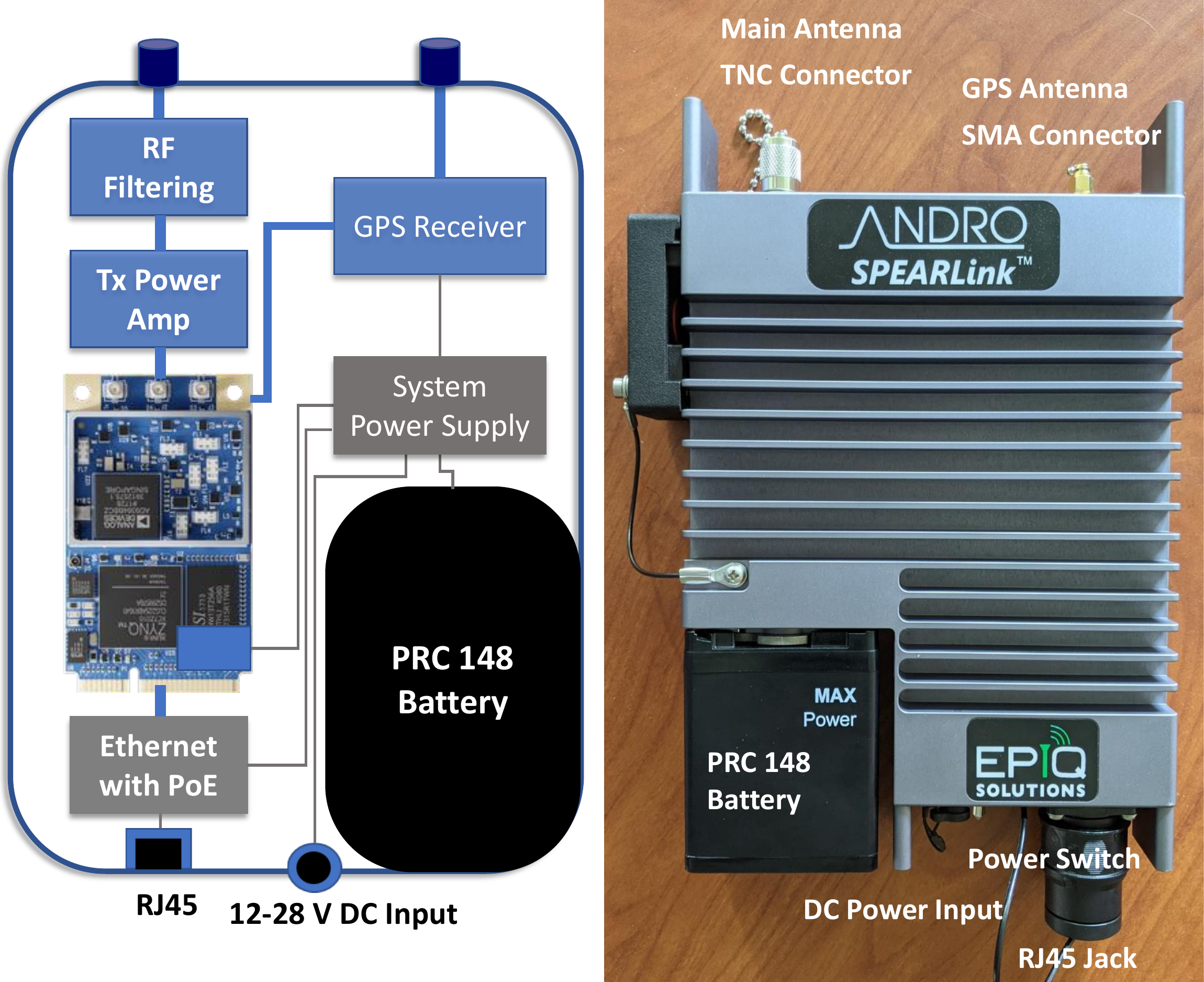} 
\caption{Transceiver Design}
\label{fig:Spearlink}
\end{minipage}
\begin{minipage}[h]{0.46 \linewidth}
\centering
\includegraphics[width=.95 \columnwidth]{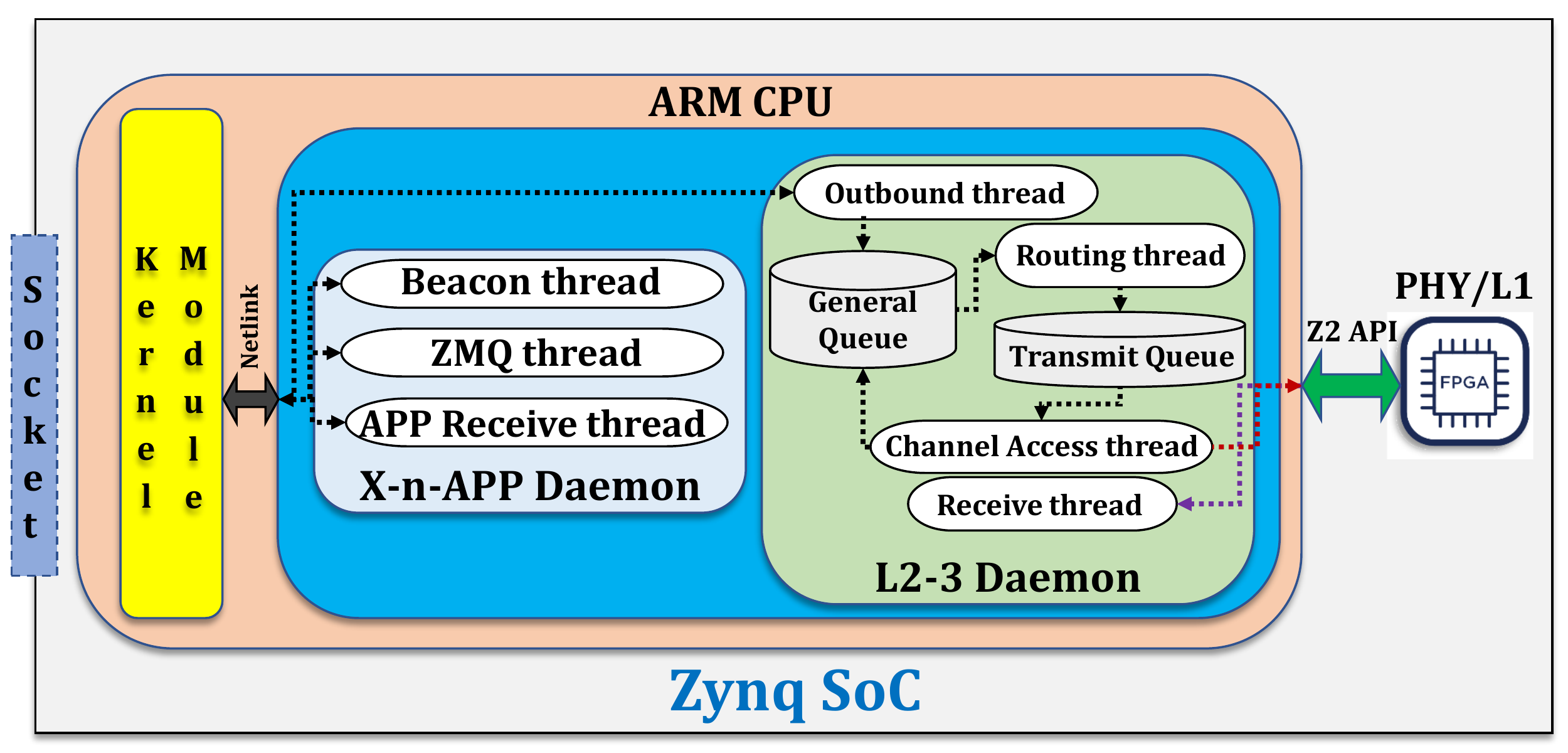} 
\caption{Software Architecture.}
\label{fig:SW}
\end{minipage}
\begin{minipage}[h]{0.27 \linewidth}
\centering
\includegraphics[width=.99 \columnwidth]{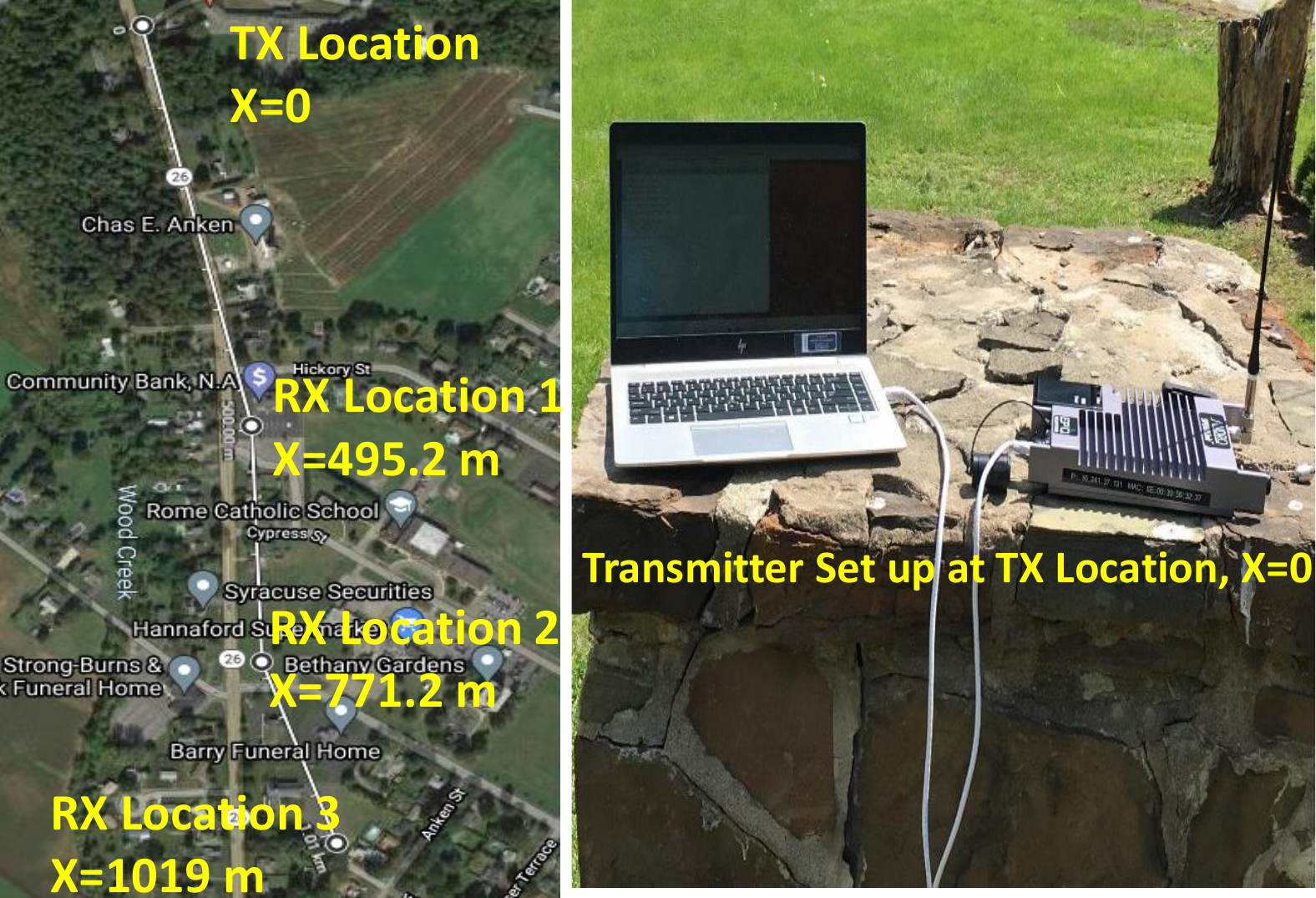} 
\caption{Range testing setup and map-view of the locations.}
\label{fig:RangeTesting}
\end{minipage}
\squeezeupp
\end{figure*}

Several hardware customizations were necessary to accomplish the objective of designing a fieldable transceiver using the COTS SDR. This includes power amplifier, filters for the frequency of interest, power supply system that is capable of supplying power from battery during standalone remote operation but could also operate from DC power source when available. To aid the implementation of various cross-layer routing techniques that use location of nodes (such as geographical routing \cite{Jagannath19ADH_HELPER}) an embedded GPS receiver was also included in the final design. The block diagram of the final transceiver and the ruggedized prototype is shown in Fig. \ref{fig:Spearlink}.

\subsection{Cross-layer Software Architecture for Embedded System}

The custom cross-layer protocol stack for the transceivers is implemented on the Zynq SoC of the Sidekiq Z2 platform. Specifically, the PHY is an IEEE802.11b implemented purely on the FPGA subsystem while the upper layers are implemented in the C/C++ language on the embedded Linux operating system of the Dual-core ARM Cortex A9 CPU of the SoC. In this section, we will detail the software architecture of the upper layers (above PHY/L1) to establish long-range mesh networking. 

Our primary goal in building the software architecture was to maintain \emph{reconfigurability}. The reconfigurability was enforced by adopting a modular design framework with defined functions for each module while being resource-efficient. We define reconfigurability as the ability to modify the protocol characteristics such as routing objective, specifics of the cross-layer information exchanged, etc.  
Essentially, designing the entire radio stack on the SoC presents a non-trivial challenge owing to the memory, computational, and latency constraints. Since an unorganized framework could add overhead from unnecessary resource utilization consequently increasing the system latency. 

The software architecture is broadly categorized into user and kernel space with daemons running in the user space as in Fig. \ref{fig:SW}. We resort to daemons each of which hosts its own threads to attain a parallelized architecture that does not perform redundant and unnecessary memory accesses. In other words, we can say the user space hosts multi-threaded processes which interface with the kernel space. However, there are certain design challenges with moving key functions to the kernel space. Kernel programming requires the most trusted operations as any bug or corruption may cause severe system crashes. Nevertheless, kernel space enjoys the benefits of low latency memory access operations. Furthermore, we also leverage the transport layer and IP headers supported by the Linux kernel. Figure \ref{fig:SW} shows a socket which is a generalized custom socket architecture for interfacing with external devices. The socket architecture is generalized to be able to reprogram and enable interfacing with a wide range of socket protocols such as ZMQ, UDP, etc.

The stack employs two daemons, namely; L2-3 and Cross-layer \& Application (X-n-APP) that interact with the kernel module via Netlink sockets. Notice that the entire stack only has two daemons since it's more efficient to use threads over processes as they share the process' resources. On the other hand, having more processes would require more computational and memory resources. The kernel module handles the transport layer, packet encapsulation, and partial IP header population prior to passing over to the L2-3 daemon. The L2-3 daemon is the cross-layer L1, L2 (MAC), and L3 (Network) module which performs CSMA/CA-based medium access and SEEK routing. Another level of cross-layer interactions occur when the L2-3 daemon acquires L1 information such as the link reliability, data rate, etc., via the Sidekiq Z2's PHY application programming interface (API). Hence, the term cross-layer as it involves interaction between L1, L2, and L3 to perform the optimized decision making. 
Due to limited space, we do not delve into the details of the cross-layer SEEK routing algorithm that is employed here. We urge interested readers to refer to our previous work \cite{Jagannath19ADH_HELPER} for detailed discussion. Therefore, we only define our utility function adopted to perform distributed optimized routing here,
$$ \mathcal{U}_{ij} = \mathcal{\eta}_{ij}\left(\frac{\max\left[\Delta\mathcal{Q}_{ij},0 \right]}{\mathtt{q}_{i}}\right) \left(\frac{d_{is}-d_{js}}{d_{is}} \right) \left( \frac{\mathtt{E}^{j}_{r}}{\mathtt{E}^{j}_{0}} \right), \forall j\in \mathbb{NB}_{i} $$

where $\eta_{ij}$ is the number of bits/Joule of transmission energy successfully transmitted from $i$ to $j$ and can be replaced by reliability when using constant power and modulation, $\mathtt{E}^{i}_{0}$ and $\mathtt{E}^{i}_{r}$ is initial and residual battery energy respectively, effective distance progressed by $i$ choosing $j$ is represented as $d_{is}-d_{js}$, and $\Delta\mathcal{Q}_{ij}$ is the differential backlog. To accomplish this objective in a \textit{distributed} manner, each node aims to maximize this value by just collecting information from its local neighbors. This information is exchanged by means of periodic beacon packets. 
The effectiveness of SEEK has been demonstrated in \cite{Jagannath19ADH_HELPER}. The emphasis of this work and contribution is not SEEK but the design and feasibility of maturing similar solutions for tactical applications. \textit{Due to the modular implementation, just by changing a few lines of code that defines the utility function, one can reconfigure the stack to execute a new optimization objective.}

The outbound DATA packets that are handed over to the L2-3 daemon from the kernel module are queued in the \textit{General Queue} awaiting the best route assignment. This is accomplished by the \textit{Outbound} thread which continuously listens for incoming packets from the kernel module. The \textit{SEEK} thread continuously performs route computation and assignment for the outbound packets in the \textit{General Queue}. Following route assignment, the packets will await their transmission opportunity in the \textit{Transmit Queue}. A third thread - \textit{Channel Access} - performs CSMA/CA awaiting a clear-to-send (CTS) from its intended next-hop/destination to dequeue the segment (set of DATA packets) from the \textit{Transmit Queue}. The segment for which the CTS was received will be forwarded to the FPGA via the L1 API for \textit{over-the-air} transmission. It must be noted that if the intended next-hop for a segment is unresponsive, the segment will be returned back to the \textit{General Queue} for rerouting. Additionally, a \textit{Receive} thread continuously monitors for incoming packets from the FPGA, processes, and responds (such as with CTS or ACK) accordingly based on received packet type. A supplementary auxiliary thread also tracks DATA timeouts. 

The X-n-APP daemon handles two tasks; (i) preparing the beacon packet with necessary information to be shared with immediate neighbors for distributed optimization, and (ii) handling messages for any application such as a graphical user interface (GUI) in this case. 
The beacon packets as well as the GUI requires location information to reflect the most recent coordinates on the GUI. We choose a efficient approach and access the GPS module from only one location - X-n-APP daemon - in the stack to avoid GPS pinging from multiple software locations. This retrieved GPS location information is subsequently populated in the beacon message. The GPS location extraction as well as beacon packet construction is carried out by the \textit{Beacon} thread. The \textit{Beacon} thread periodically sends this beacon messages to the lower layers for outbound transmission where the remaining fields such as residual battery and current buffer backlog are updated. Here, we note that unlike the DATA, the beacon packet is treated as control packet and are not queued in the general or transmit queues of the L2-3 daemon rather it is directly sent to the L1 FPGA for transmission.

The X-n-APP daemon also functions as a generalized application daemon for direct interfacing with an application such as a GUI. Both \textit{ZMQ} and \textit{APP receive} threads are designed to support such applications and hence can be customized based on the application at hand. Since the use case of this particular transceiver is remote deployment in test and evaluation ranges, a GUI application is developed to interface with the transceiver at a central command and control location. We emphasize that the remote monitoring and configuration (parameters like data rates, frequency etc.) capability enabled by the GUI will ease the operator load by alleviating the need to physically travel to the deployed locations. However, we do not elaborate on the GUI design or its implementation aspects as it is beyond the scope of this article rather state that the daemon is capable of handling any such applications. 

We further reemphasize that the software-defined stack is designed to be reconfigurable to modify the cross-layer routing as the requirements evolve in the future or depending on the desired network application. It is noteworthy that the utility function discussed above is representative of one such example of the algorithm where the network application desires energy-aware routing. This specific choice was dictated by the customer's requirement.\textit{ The broad impact of this work is the reconfigurable nature of the software-defined stack such that it keeps evolving to meet the future requirements rapidly.}

\section{Outdoor Experiments and Results}
\label{Sec:Resutls}


In this section, we performed extensive outdoor experiments to evaluate the performance of the implemented software modules in a realistic environment. Since the proposed solution is implemented on SDR, it can be extended to any frequency of interest supported by the target hardware. To accomplish the experiments, we received a temporary experimental license from FCC to utilize the 430 MHz frequency within a geographical area. 
It is important to point out that the purpose of the experiments is to evaluate the performance of implemented cross-layer optimized algorithms on general purpose processor (GPP) supported by a FPGA-based PHY layer for outdoor deployment. The novelty and utility of the SEEK algorithm itself has been established in previous work \cite{Jagannath19ADH_HELPER}. 
To the best of our knowledge, this is the first time such a comprehensive evaluation has been undertaken using deployment-ready cross-layer optimized embedded SDRs. The default parameters are listed in the Table \ref{tab:para} unless otherwise specified.

\begin{table}[t!]
\small
    \centering
    \caption{Parameters Of Outdoor evaluation\label{tab:para}}
    \squeezeupp
    \begin{tcolorbox}[tab2,tabularx={|p{3 cm}||p{5 cm}}]
      \textbf{Parameters}   & \textbf{Values}  \\ \hline\hline
      Data rates   & 1, 2, 5.5, 11 Mbps   \\ \hline
    Transmit Power & 1-3.5 W (based on attenuation)   \\ \hline
    Center Frequency & 430 MHz \\ \hline
    Bandwidth & 22 MHz \\ \hline
    Payload Size   & 1000 Bytes   \\ \hline
    Segment Size   & 32 Packets   \\ \hline
    \end{tcolorbox}{}
    \squeezeupp
\end{table}

\subsection{Transmission Range Evaluation}

The target deployment scenario for the device is remote test and evaluation sites, hence, the design goal for transmission range was to achieve up to $1$ km. The setup of the transmitter and the key testing locations from our range testing experiments is shown in Fig. \ref{fig:RangeTesting}. 
For each reported value in the Table. \ref{tab:Range_Test}, the results were averaged over 10,000 packets transmitted in each run. \textit{We define \textbf{reliability} as the percentage of packets received with respect to packets sent.}

\begin{table}[h!]
\small
    \centering
    \caption{Link reliability for varying transmission ranges}
    \squeezeup
    \label{tab:Range_Test}
    \begin{tcolorbox}[tab2,tabularx={|p{2 cm}||p{2.7 cm}||p{2.5 cm}}]
      \textbf{Data Rate}   & \textbf{Distance} &\textbf{Reliability}    \\ \hline \hline
    \multirow{4}{*}{2 Mbps}  & 495.2 m  &   99.9\%   \\ \cline{2-3}
      & 771.2 m  &  99.77\%   \\ \cline{2-3}
      & 1019 m  &   98.08\%   \\ \cline{2-3}
\hline
    \multirow{4}{*}{5.5 Mbps}  & 495.2 m  &  99.62\%   \\ \cline{2-3}
      & 771.2 m  &  96.03\%   \\ \cline{2-3}
      & 1019 m  &  97.16\%   \\ \cline{2-3}
\hline
      \multirow{4}{*}{11 Mbps}  & 495.2 m  &  85.28\%   \\ \cline{2-3}
      & 771.2 m  &  31.56\%   \\ \cline{2-3}
      & 1019 m  &  13.9\%   \\ \cline{2-3}
    \end{tcolorbox}{}
    \squeezeup
\end{table}

\begin{figure*}[h!]
\begin{minipage}[h]{0.33 \linewidth}
\centering
\includegraphics[width=.95 \columnwidth]{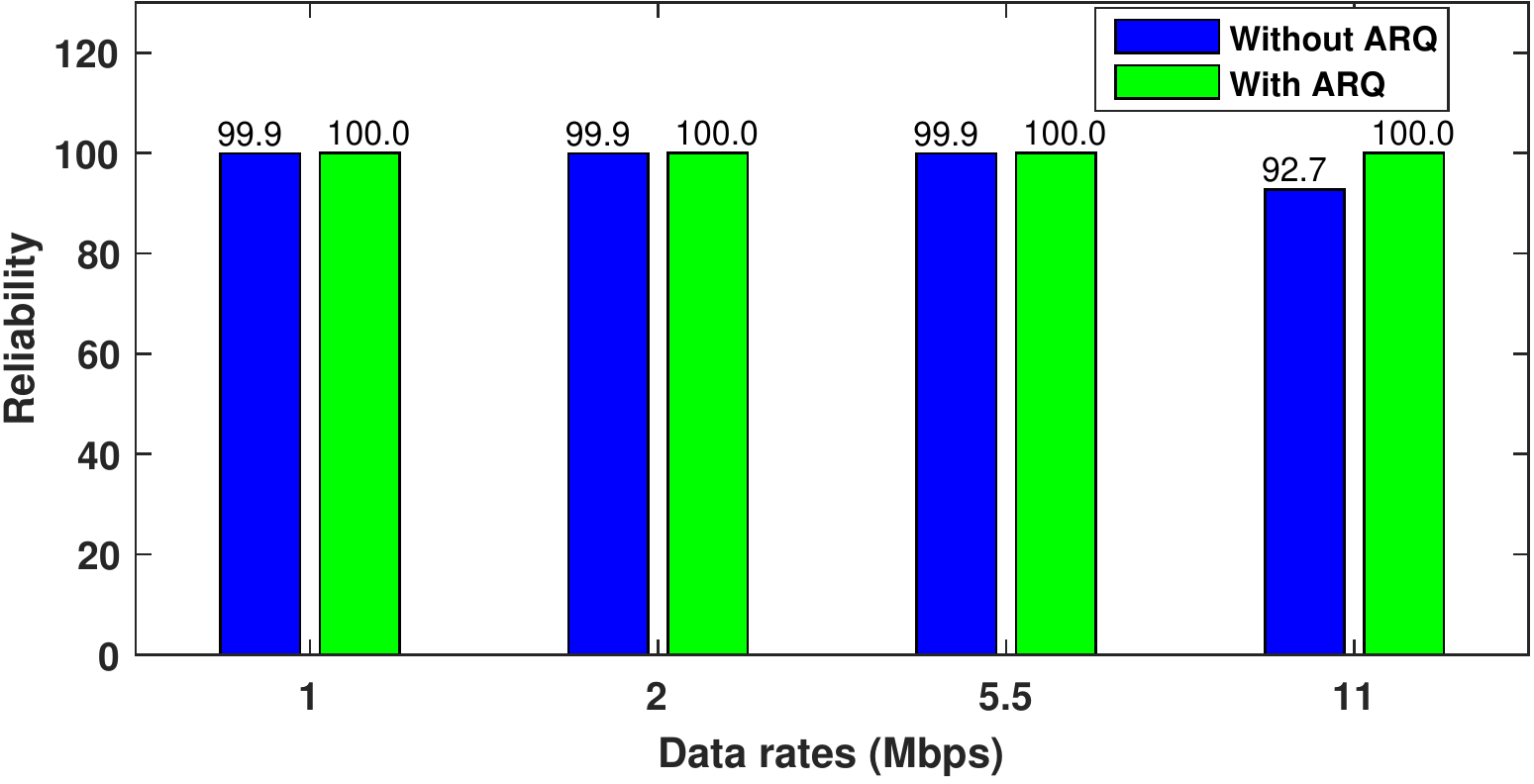} 
\caption{Reliability vs data rates}
\label{fig:RL_peer}
\end{minipage}
\begin{minipage}[h]{0.33 \linewidth}
\centering
\includegraphics[width=.95 \columnwidth]{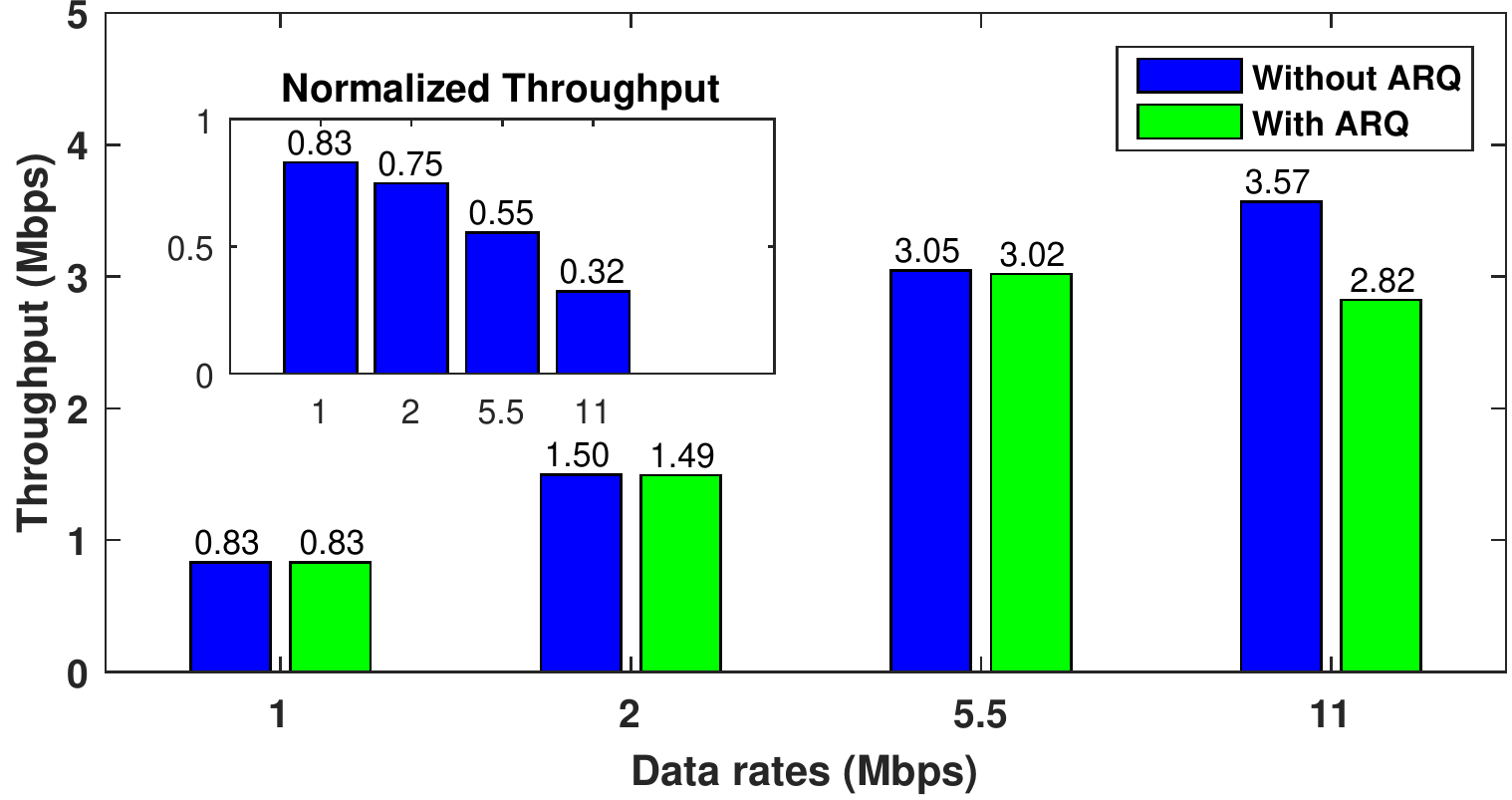} 
\caption{Throughput vs data rates}
\label{fig:Th_peer}
\end{minipage}
\begin{minipage}[h]{0.33 \linewidth}
\centering
\includegraphics[width=.95 \columnwidth]{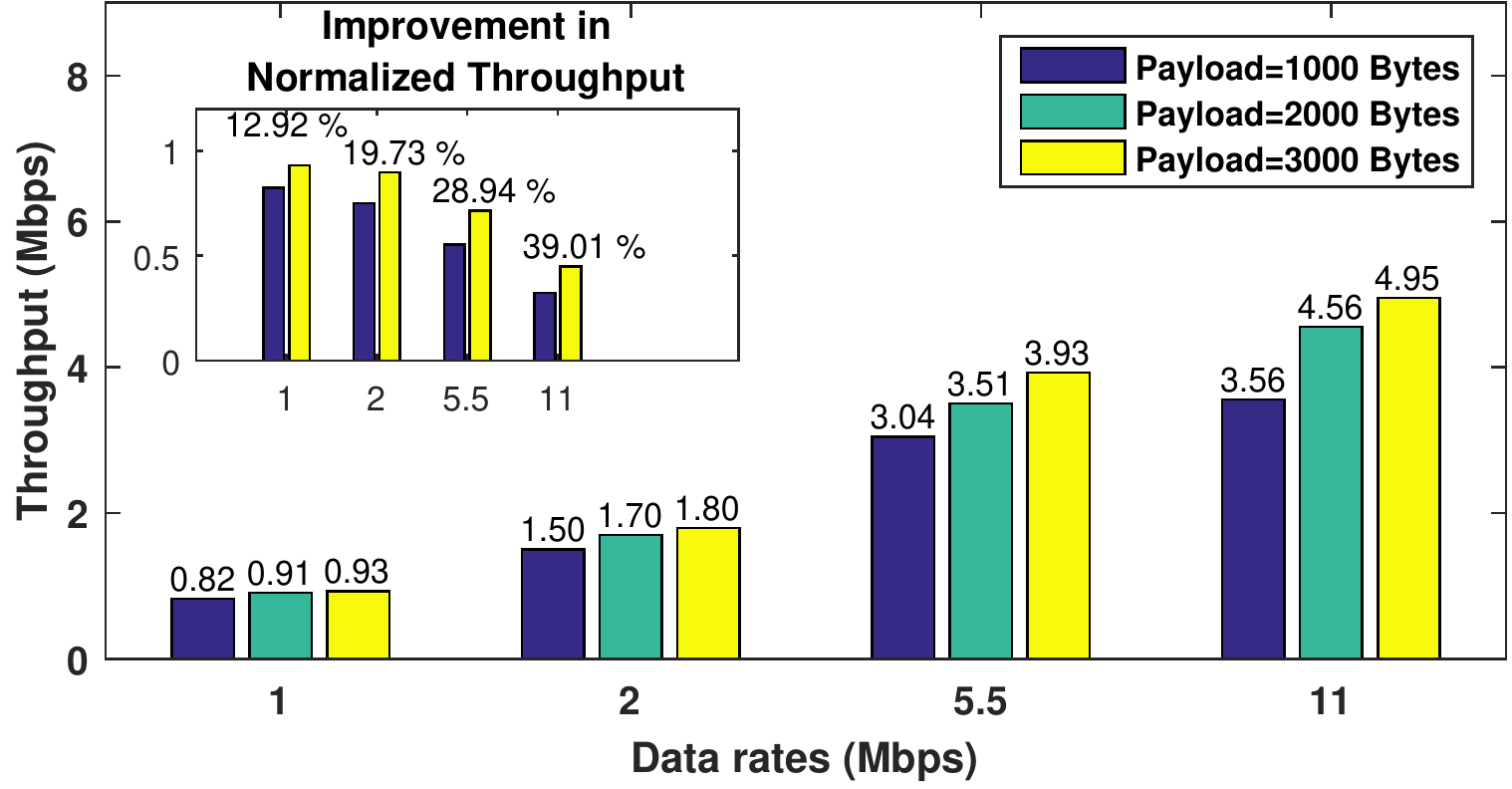} 
\caption{Throughput for varying payload}
\label{fig:Th_payload}
\end{minipage}
\begin{minipage}[h]{0.24 \linewidth}
\centering
\includegraphics[width=.99 \columnwidth]{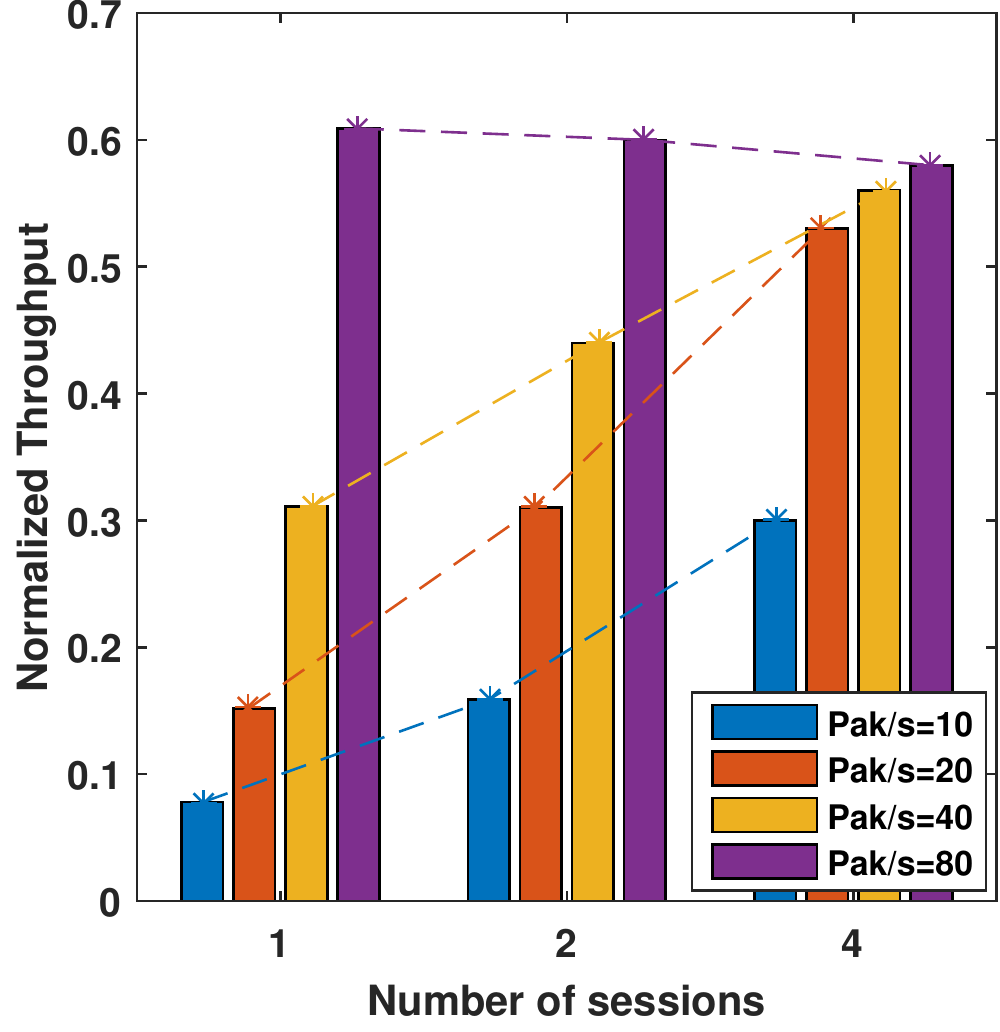} 
\caption{10-node Experiment}
\label{fig:Network}
\end{minipage}
\begin{minipage}[h]{0.45 \linewidth}
\centering
\includegraphics[width=.95 \columnwidth]{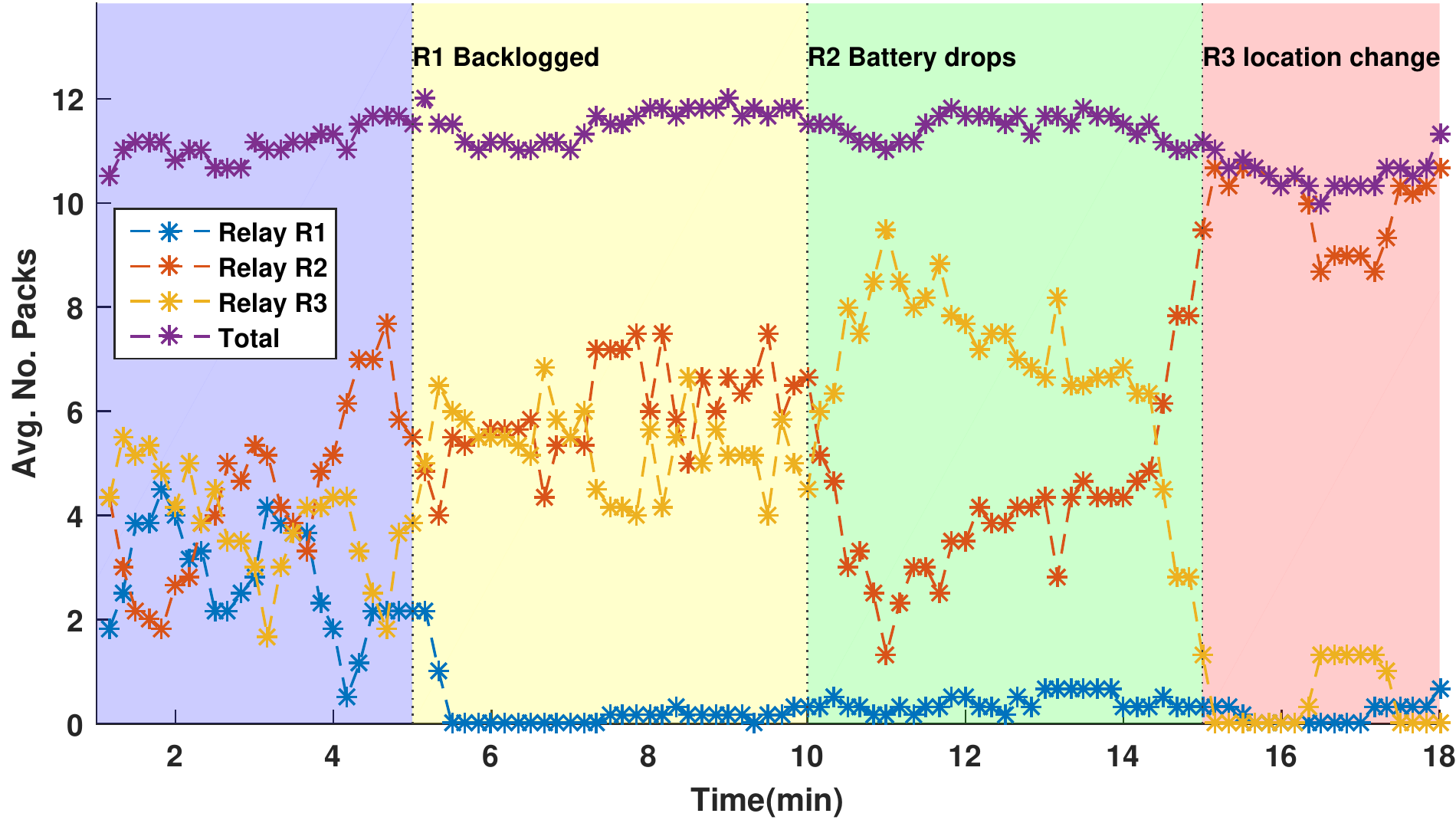} 
\caption{5-node Experiment (dynamic routing)}
\label{fig:route}
\end{minipage}
\begin{minipage}[h]{0.3 \linewidth}
\centering
\includegraphics[width=.99 \columnwidth]{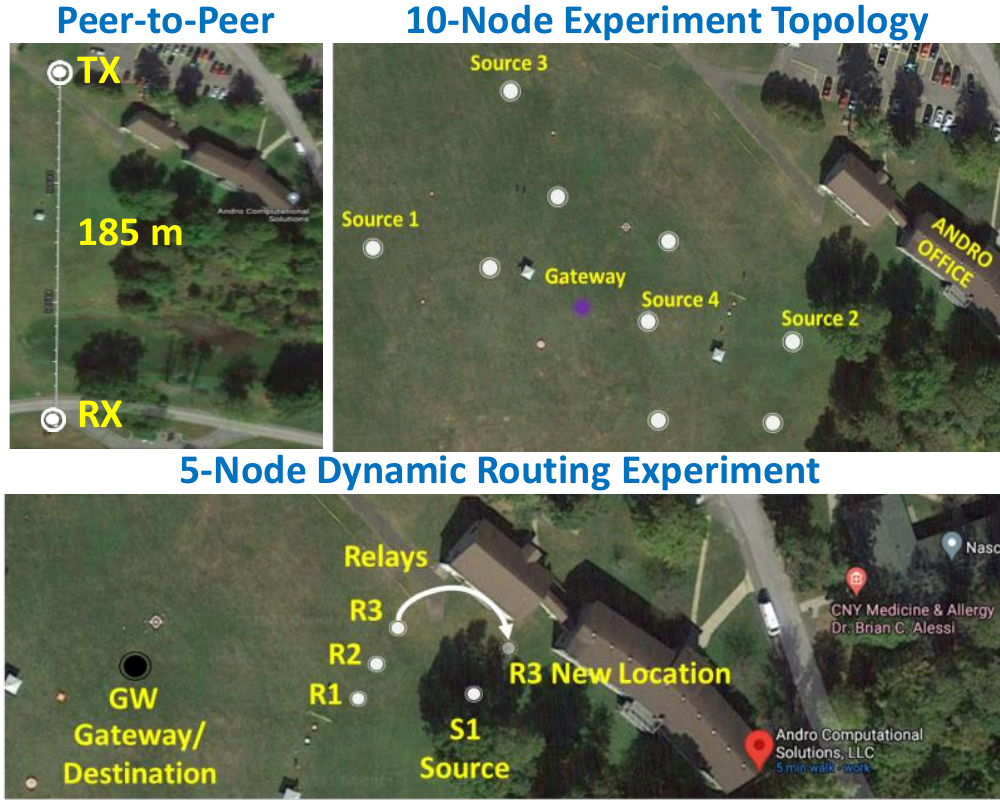} 
\caption{Outdoor Topologies}
\label{fig:TopologyMaps}
\end{minipage}
\squeezeupp
\end{figure*}

Table. \ref{tab:Range_Test} shows that the performance was consistent for both $2$ Mbps and $5.5$ Mbps up to $1019$ m but the performance of $11$ Mbps degraded sooner (1 Mbps eliminated due to limited space). Due to the FCC license restrictions and urban environment, the next feasible point was at $1942$ m from the transmitter at which point no packets were received. The performance degradation of $11$ Mbps was attributed to sensitivity to lower signal-to-noise-ratio and multi-path propagation effects. Further in-lab cabled analysis revealed there is room for improvement in sensitivity (~15 dB) in the future iterations of the FPGA PHY. 
Overall, the experimentation was successful as we hit our target distance of 1 km even at $5.5$ Mbps. 

\subsection{Peer-to-Peer Experiments}

As described earlier, the decision to implement the PHY on FPGA and the rest of the protocol stack was to achieve the best combination of high data rates and the required flexibility to implementing cross-layer routing protocols. Yet, it is important to study the delays and associated overhead it takes to make the necessary calls between the GPP and FPGA to support packet transmission and how it impacts the data rates. To accomplish this, we perform outdoor peer-to-peer testing for varying data rates and payload sizes to analyze the impact.

First, in Fig. \ref{fig:RL_peer}, we study the reliability at the different data rates with and without automatic repeat request (ARQ) enabled at the MAC layer. It can be clearly seen that the transceiver has high reliability close to $100\%$ for 1, 2, and 5.5 Mbps even without ARQ enabled. At 11 Mbps the reliability decreases to $92\%$ without ARQ and returns to $100\%$ with ARQ with the trade-off of throughput. Figure \ref{fig:Th_peer} depicts how the peer-to-peer throughput varies at different PHY data rates 1, 2, 5.5, 11 Mbps. \textit{The \textbf{throughput} in this case only considered the payload of the packets and does not include the headers or the control packets}. The throughput considered here is also referred to as goodput in some literature. The figure also depicts the \textit{\textbf{normalized throughput} which is defined as the ratio between throughput and data rate}. The design choice provides solid performance at 1 and 2 Mbps especially considering it depicts the goodput achieved. The impact of the overhead becomes visible as the data rate increases to 5.5 and 11 Mbps. This hypothesis is further substantiated in our experiments that show an increase in normalized throughput with the increase in payload as shown in Fig. \ref{fig:Th_payload}. The normalized throughput increased by up to $39\%$ for 11 Mbps when the payload was increased from $1000$ Bytes to $3000$ Bytes.     

\subsection{10-Node Network Experiments - Network Capacity}

In this experiment, we evaluate how the network handles increasing number of sessions and source (packet generation) rate. The 10-node topology is shown in Fig. \ref{fig:TopologyMaps}. The experiments were conducted at $1$ Mbps of data rate, the source rate was varied from $[10, 20, 40, 80]$ packets/s, and the number of sessions (independent sources generating traffic to the gateway node) were set to $1$, $2$ and $4$. This is motivated by a typical use case for our intended deployment scenario. As expected, Fig. \ref{fig:Network} demonstrates that the normalized throughput increases and gets saturated both when the source rate increases and the number of sessions increases. There is no drop in performance even when there are multiple sessions ($4$) with high source rates ($80$) and the normalized throughput is maintained at the saturated value ($\sim0.6$). This shows the efficiency of the network in handling multiple traffics and session rates as expected even when the traffic increases. 

\subsection{5-Node Network Experiment - Dynamic Routing}

For timely evaluation, we designed a specific experiment such that we can determine the effectiveness of the implementation of SEEK algorithm to dynamic changes in the network. To accomplish this, we consider a topology shown in Fig. \ref{fig:TopologyMaps}. The source \textit{S1} continuously transmits packets destined for the gateway node \textit{GW}. \textit{S1} chooses the appropriate relay among \textit{R1}, \textit{R2}, and \textit{R3} based on SEEK algorithm. We plot the average packets received from each relay node every $10$ s interval. The values are averaged using a moving window of $60$ s duration to get smoother curves.

In the first part of Fig. \ref{fig:route} (blue shade), in a uniform setting, the traffic is evenly distributed. Next, we increase the backlog of the \textit{R1} to emulate a congestion scenario at a node. As it can be seen (yellow shade), \textit{S1} learns to avoid \textit{R1} and routes the packets through \textit{R2} and \textit{R3}. Next, the residual energy of \textit{R2} is rapidly dropped by removing the DC power supply and the residual energy in the battery was close to $10\%$. As it can be seen, \textit{S1} recognizes this in the next part (yellow shade) and prefers \textit{R3} so that lifetime of the network can be extended. Finally, we now move the \textit{R3} node behind the source \textit{S1} as shown in Fig. \ref{fig:TopologyMaps} such that it is no more a feasible next hop. In this situation, with \textit{R1} congested, \textit{R3} is no more the possible choice for forward progress, \textit{S1} must return to \textit{R2} with a lower battery since that is the only possible choice to relay the packets to the destination. \textit{It is also interesting to see that even with all these changes, the total packets/s at the \textit{GW} remains very stable demonstrating the rapid adaptability of the cross-layer optimized nodes.} In this experiment, we have demonstrated the gamut of routing decisions the nodes can make in a distributed manner by just gathering information from its immediate neighbors demonstrating the effectiveness of the cross-layer optimized routing using an intuitively representative experiment.   
\squeezeu

\section{Conclusion and Future work}

This article introduces the first known cross-layer optimized transceiver that has been designed, developed, and matured to provide high throughput and reliability by implementing the protocol stack (except PHY) on an embedded ARM processor. We have demonstrated through extensive outdoor experiments the capability, range, and demonstrated dynamic routing under varying network conditions. The modular design allows reconfiguration of the cross-layer protocol stack to rapidly customize components like, information flow, cross-layer interactions, and optimization objective itself by changing a few lines of codes. We hope this successful demonstration of cross-layer optimized embedded transceivers will provide directions for future cross-layer optimized solutions to benefit tactical and commercial applications. 
\squeezeu

\section*{Acknowledgment}

Authors would like to thank John Orlando, Jeff Porter of Epiq Solutions for their support, Raymond Shaw of Spectrum Bullpen for help with SSRA, and Dan O' Connor of ANDRO for help during the outdoor testing. 

\ifCLASSOPTIONcaptionsoff
  \newpage
\fi



%
\bibliographystyle{IEEEtran}	
\bibliography{Andro1}
\squeezeupp
\vskip -2.5\baselineskip

\end{document}


\title{\textbf{Cover Letter}}
\maketitle
Due to the 15-count reference limit in the magazine article, all of the related works are not cited in the manuscript. Instead, we have only cited the most relevant ones. This document is provided to serve as a supplementary information to highlight the key contributions of the article and its distinguishing factors with respect to prior research in this field to assist the reviewers and editors. 

There have already been some works on state-of-the-art deep learning approaches as well as future 6G networks. All presented in this document is compared and contrasted here for clarity. 
\begin{enumerate}
    \item The work in [1] studies the key challenges and potential physical layer solutions for 6G. This work talks about algorithmic physical layer approaches and briefly mention deep neural network as a key 6G enabler. However, they do not mention how these physical layer techniques can be realized for an edge computing platform to facilitate the key concept of edge learning in 6G.
    
    \item The authors of [2] present a vision of the 6G network in terms of the various use cases, challenges, and key enabling technologies. However, the focus here is on an overall 6G network and not specifically the physical layer solutions. The authors however mention AI as a pervasive enabler in realizing 6G.
    
    \item In [3], an intelligent edge concept for wireless communication is elaborated. The learning-driven radio resource management and signal encoding problem is studied in depth. However, the study on complexity of the approaches and their implementation platforms (edge computing devices) is lacking. 
    
    \item The authors of [4] presents the various 6G network use cases and enabling technologies. However, the authors only briefly mention integrating AI in the 6G network towards achieving autonomous operation and do not go in-depth into their implementation challenges. 
    
    \item The authors of [5] discusses the 6G architecture and a few promising technologies to realize the 6G requirements. Authors mention the various AI techniques that will serve as key enablers for the intelligent 6G architecture while succinctly mentioning model-driven deep learning approaches as one of the candidate AI techniques.
    
    \item In [6], traditional deep learning for the optical wireless communication system is studied. Specifically, a convolutional autoencoder for image sensor communication is presented and evaluated in simulations. However, the computational complexity and processing delay of such an approach will be very high owing to the maximum likelihood complexity of the decoding layer alone. An in-depth study on the implementation challenges is lacking in this work.
    
    \item The authors of [7] presents the architecture of 6G networks while discussing the concept of intelligent radio. They also briefly mention the importance of an intelligent physical layer in realizing the 6G communication requirements. However, their work does not present the candidate physical layer approaches that will realize the intelligent physical layer concept.
    
    \item In [8], the authors present a detailed study of the traditional machine learning techniques that have been applied to solving wireless communication problems. However, their complexity, latency, and reliability from a hardware-efficient implementation standpoint are lacking in this work.
\end{enumerate}

 Indeed, irrespective of the AI-based upper layer schemes, the communication capabilities in terms of reliability, processing delay, latency, and complexity of a radio is bottlenecked by its underlying physical layer. In this work, we focus on how deep unfolded approaches that integrates deep learning with principled algorithms advances the communication capabilities in realizing a hardware-efficient intelligent physical layer. We aim to motivate the reader in understanding how computational intensive deep learning approaches may not be the solution to achieve the 6G communication requirements. Instead, integrating the powerful learning capability of deep learning with algorithmic knowledge will relieve the computational burden as well as improve the robustness. Therefore, this is the first work that studies the potential intelligent hardware-efficient physical layer solutions that would serve as key enablers to realize the 6G AI radio concept.
\\
\\
\textbf{Related works}

[1] P. Yang, Y. Xiao, M. Xiao and S. Li, "6G Wireless Communications: Vision and Potential Techniques," in IEEE Network, vol. 33, no. 4, pp. 70-75, July/August 2019.
\\

[2] F. Tariq, M. R. A. Khandaker, K. Wong, M. A. Imran, M. Bennis and M. Debbah, "A Speculative Study on 6G," in IEEE Wireless Communications, vol. 27, no. 4, pp. 118-125, August 2020.
\\

[3] G. Zhu, D. Liu, Y. Du, C. You, J. Zhang and K. Huang, "Toward an Intelligent Edge: Wireless Communication Meets Machine Learning," in IEEE Communications Magazine, vol. 58, no. 1, pp. 19-25, January 2020.
\\

[4] M. Giordani, M. Polese, M. Mezzavilla, S. Rangan and M. Zorzi, "Toward 6G Networks: Use Cases and Technologies," in IEEE Communications Magazine, vol. 58, no. 3, pp. 55-61, March 2020.
\\

[5] Y. Zhao, J. Zhao, W. Zhai, S. Sun, D. Niyato, and K-Y. Lam, “A survey of 6g wireless communications: Emerging technologies,” ArXiv, vol. abs/2004.08549, 2020.
\\

[6] H. Lee, S. H. Lee, T. Q. S. Quek and I. Lee, "Deep Learning Framework for Wireless Systems: Applications to Optical Wireless Communications," in IEEE Communications Magazine, vol. 57, no. 3, pp. 35-41, March 2019.
\\


[7] K. B. Letaief, W. Chen, Y. Shi, J. Zhang and Y. A. Zhang, "The Roadmap to 6G: AI Empowered Wireless Networks," in IEEE Communications Magazine, vol. 57, no. 8, pp. 84-90, August 2019.
\\

[8] J. Wang, C. Jiang, H. Zhang, Y. Ren, K. Chen, and L. Hanzo, “Thirty years of machine learning: The road to pareto-optimal next-generation wireless networks,” ArXiv, vol. abs/1902.01946, 2019.


\begin{appendices}
\section{Formulation of optimization problem}\label{Ap:opt}

The global objective of the optimization problem is to find the optimal global vectors $\mathbf{R}$, $\mathbf{F}$ and $\mathbf{P}$ that maximizes the sum of the network utilities, under the constraints of power and $BER$. The formulation of the optimization problem is as follows,

\begin{align}
\mathcal{P}_1\!:\textup{Given}\!&: \mathcal{G(U,E)},\;\;P^{Bgt},\;\; Q_i^s,\;\; BER_{\mathcal{SU}},\;\;BER_{\mathcal{PU}}\notag \\
				\textup{Find}\!&: \mathbf{R},\;\; \mathbf{F},\;\; \mathbf{P},\;\; \mathbf{A}\notag \\
				\textup{Maximize}\!&: \sum_{i \in \mathcal{SU}} \sum_{j \in \mathcal{SU}/i} U_{ij}(a_i^s(t))\\
				\textup{subject\; to}\!&:\notag \\
				& \sum_{s \in S} r_{ij}^s \leq C_{ij}, \forall i \in \mathcal{SU},\; \forall j \in \mathcal{SU} \label{constr:capacity_P1} \\ 
				& SINR_k \geq SINR_{\mathcal{PU}}^{th}(BER_{\mathcal{PU}}), \forall k \in \mathcal{PU}, \forall f \label{constr:PU_P1} \\ 
				& SINR_l \geq SINR_{\mathcal{SU}}^{th}(BER_{\mathcal{SU}}), \forall l \in \mathcal{SU}, \forall f \label{constr:SU_P1} \\ 
				& \sum_{f \in[f_i, f_{i+\Delta f_i}]} P_i(f) \leq P_i^{Bgt}, \forall i \in \mathcal{SU} \label{constr:Power_P1}
\end{align}

In the above formulation, the objective is to maximize the network utility of all the active links. The constraint (\ref{constr:capacity_P1}) restricts the total amount of traffic in link $(i,j)$ to be lower than or equal to the physical link capacity. Constraint (\ref{constr:PU_P1}) and (\ref{constr:SU_P1}) imposes that any transmission by secondary user should guarantee the required $BER$ for the active primary users and secondary user respectively. Finally, $P_i^{Bgt}$ is the instantaneous power available at the cognitive radio.
\end{appendices}





\ifCLASSOPTIONcaptionsoff
  \newpage
\fi